# Accelerated Discovery of Two-Dimensional Optoelectronic Octahedral Oxyhalides via High-Throughput *Ab Initio* Calculations and Machine Learning

Xing-Yu Ma[1], James P. Lewis[2,3,4], Qing-Bo Yan[5*], Gang Su[1,6*]

[1]School of Physical Sciences, University of Chinese Academy of Sciences, Beijing 100049, China.

[2]Department of Physics and Astronomy, West Virginia University, Morgantown, West Virginia 26506-6315, United States

[3]State Key Laboratory of Coal Conversion, Institute of Coal Chemistry, Chinese Academy of Sciences, Taiyuan, Shanxi 030001, China

[4]Beijing Advanced Innovation Center for Materials Genome Engineering, Beijing Information S&T University, Beijing 101400, China

[5]Center of Materials Science and Optoelectronics Engineering, College of Materials Science and Optoelectronic Technology, University of Chinese Academy of Sciences, Beijing 100049, China.

[6]Kavli Institute for Theoretical Sciences, and CAS Center of Excellence in Topological Quantum Computation, University of Chinese Academy of Sciences, Beijing 100190, China

**ABSTRACT: Traditional trial-and-error methods are obstacles for large-scale searching of new optoelectronic materials. Here, we introduce a method combining high-throughput *ab initio* calculations and machine-learning approaches to predict two-dimensional octahedral oxyhalides with improved optoelectronic properties. We develop an effective machine-learning model based on an expansive dataset generated from density functional calculations including the geometric and electronic properties of 300 two-dimensional octahedral oxyhalides. Our model accelerates the screening of potential optoelectronic materials of 5,000 two-dimensional octahedral oxyhalides. The distorted stacked octahedral factors proposed in our model play essential roles in the machine-learning prediction. Several potential two-dimensional optoelectronic**

**octahedral oxyhalides with moderate band gaps, high electron mobilities, and ultrahigh absorbance coefficients are successfully hypothesized.**

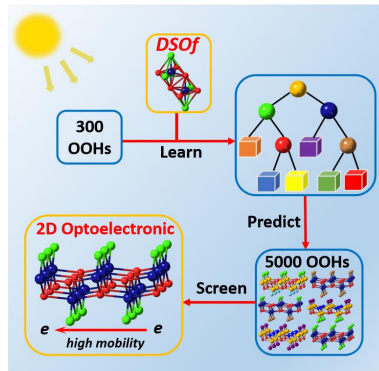

Functional materials are the cornerstone of modern industry, and the exploration of materials with excellent properties represents the frontier of materials science. Overcoming the difficulty of traditional trial-and-error methodologies requires high-throughput *ab initio* calculations based on density functional theory (DFT) applied to large-scale searching of appropriate materials.[1] Examples include screening two-dimensional (2D) topological materials[2] from Materials Project,[3] or searching easily exfoliable 2D materials[4] from Inorganic Crystal Structure Database (ICSD)[5] or the Crystallographic Open Database,[6] or exploring 2D piezoelectric materials[7] from ICSD, *etc*. However, there are only very few materials with expected properties hiding among thousands or even more candidates in materials databases. Thus, most of the expensive DFT calculations in searching for proper materials are performed under a very low efficiency. Recently, rapid development of the artificial intelligence technology, especially the machine-learning methodologies, has spurred intensive applications in materials and chemical sciences,[8-11] such as designing functional materials without inversion symmetry,[12] predicting properties of inorganic crystals,[13] discovery of organic-inorganic perovskites, *etc*.[14] These advances illustrate that machine-learning techniques help not only to establish quantitative materials structure-property relationships but also to accelerate the discovery of new functional materials.

Since the discovery of graphene[15] in 2004, 2D materials have received much attention not only for being the thinnest crystalline solids but also for potential applications in future electronics and optoelectronics.[16, 17] Various types of 2D materials have also been reported, including transition metal dichalcogenides (TMDCs),[18] hexagonal boron nitride,[19] group IV monochalcogenides[20] and phosphorene,[21] etc. However, the zero band gap of graphene is the obvious obstacle for applications in transistors;[22] the low carrier mobility limits practical applications of TMDCs monolayers,[23, 24] while poor stability of phosphorene brings about sharp degradation of performance.[25, 26] Therefore, there is a growing interest in discovering new 2D semiconductors with excellent optoelectronic properties including moderate band gap, high mobility, and prominent absorption and good stability. Recently, single-layer semiconductors CrOX (X=Cl/Br) belonging to octahedral oxyhalides (OOHs) have been proposed,[27] which could be easily prepared from their bulk forms that have layered structures with weak van der Waals interactions and small exfoliation energy.[28-30] In this family of 2D materials, excellent optoelectronic materials potentially exist, but to the best of our knowledge, they have not been investigated previously.

In this work, we present a method combining high-throughput *ab initio* calculations and machine learning to explore 2D OOHs family to find proper materials with excellent optoelectronic properties. Our proposed workflow is as follows. First, we calculate the geometric and electronic properties of 300 different OOHs from DFT calculations to compose an input dataset for machine-learning model training. Second, we train many different machine-learning algorithms from this dataset to evaluate the best machine-learning algorithm. Third, the most effective machine-learning model is utilized to predict the electronic properties of 5000 OOHs, and thereby, accelerate the screening of possible optoelectronic materials. We screen OOHs choosing criteria such as non-toxicity, low cost, dynamical stability, high mobility, and enhanced optical absorption. Finally, we propose several OOHs with excellent optoelectronic properties which may have promising applications in areas like high-speed electronic devices or optoelectronic devices.

As shown in Figure 1(a), 2D OOHs ($A_2B_2X_2$, B=O/S/Se/Te, X=F/Cl/Br/I) crystallize in an orthorhombic structure and consists of the undulant $A_2B_2$ layers sandwiched by the halogen atoms. The structures can also be viewed as up-down alerted stacking of distorted octahedron centered by A atom and bounded by the chalcogens B1/B2 and halogens X1/X2 (Figure 1(b)). In particular, the halogens X1 and X2 are stacked up and down along the $A_2B_2$ layers, while the chalcogens B1 and B2 are staggered on the up and down of the $A_2B_2$ layers. Note that when X1 and X2 or B1 and B2 are different, a Janus-like structure will be formed. It has been reported that Janus TMDC MoSSe has been successfully synthesized in lab[31] and heterostructures of Janus MoSSe and WSSe give rise to novel and unexpected optoelectronic properties.[32] The Janus structure may lead to enhanced photoelectric performance. We generate diverse OOHs in our practice by replacing the A atom by 53 elements across the periodic table, while halogens (F, Cl, Br, I) and chalcogens (O, S, Se, Te) are collected as potential replacements for X1/X2 and B1/B2 alternately occupying vertices of two distorted octahedrons, respectively (Figure 1(b)). Stoichiometrically equivalent structures are excluded, for instance, X1=F, X2=Cl and X1=Cl, X2=F are viewed as the same combination. In our approach, 5300 different OOHs are considered. A thorough DFT investigation on all these materials is obviously time-consuming and very expensive. To remedy this, we use the machine-learning technique combing with high-throughput *ab initio* calculations to accelerate the searching and screening of the OOHs materials.

The procedure utilized to discover proper photoelectric materials in OOHs is illustrated in Fig. 1(c). The prediction project consists of four components: input data, machine-learning model, machine-learning prediction, and preliminary screening as shown in the left panel of Fig. 1(c). We generate an input dataset from DFT calculations on the geometric and electronic properties of 300 2D OOHs, each of which is described by 62 initial features (descriptors). The 300 OOHs were selected randomly from all 5300 OOHs to guarantee that the training set covered the parameter space of all data. Based on the input dataset, we train and test different machine-learning models and find gradient boosted regression (GBR) [33] model is the best choice. In addition, we use a feature reduction algorithm to remove redundant features and obtain a minimum model complexity of 26 descriptors, which are ranked according to the degree of importance. With this machine-learning model, we predict the electronic properties of remaining 5000 OOHs and screen out 411 OOHs with the criteria of band gaps between 0.9 and 1.6 eV.[34, 35] Considering that toxicity and expensiveness are always disadvantageous towards broader applications, we further exclude materials containing non-toxic and low-cost metal elements from above 411 OOHs. As a result, a remaining 73 OOHs are finalized in the selection process. As shown in the right panel of Fig. 1(c), these 73 screened OOHs were systematically checked by first-principles calculations to generate a number of expected materials with excellent optoelectronic performance. We test the dynamical stability of the 73

OOHs by using density functional perturbation theory (DFPT),[36] and 20 OOHs exhibit the proper stability. We further calculate the effective mass of charge carriers and carrier mobility based on the deformation potential theory[37], finding that six OOHs have proper band gaps and high carrier mobilities. Finally, using $G_0W_0$ and Bethe-Salpeter equation (BSE) methods,[38, 39] three OOHs ($Bi_2Se_2Br_2$, $Bi_2Se_2BrI$, and $Bi_2Se_2I_2$) are revealed to have ultrahigh light-absorption coefficients. Thus, they are promising candidates for possible applications in optoelectronic devices. Details of our calculations are found in Supplementary Information (Table S3, S6; Figures S6, S8, S13, and S14).

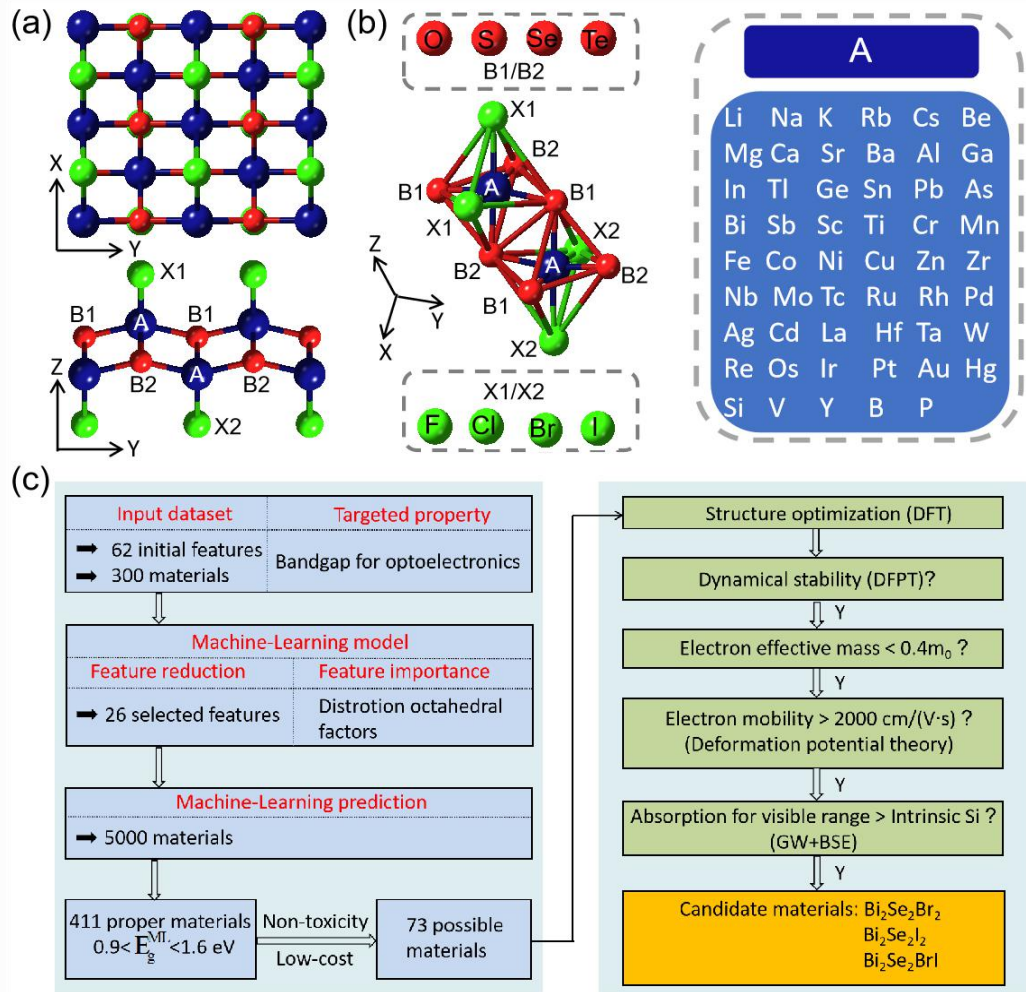

Figure 1. Schematic structure of octahedral oxyhalides (OOHs) and the procedure of prediction and screening. (a) Top and side views of 2×2 unit cell of a typical OOH structure. (b) Distorted stacking octahedrons centered by A atom and surrounded by the chalcogens B1/B2 and halogens X1/X2. The combination of 53 elements (A-site), 4 representative chalcogens (B1/B2) and 4 typical halogens (X1/X2) constitutes all 5300 OOHs candidates. (c) The left panel represents the prediction project consisting of four components: Input dataset, machine-learning model, machine-learning prediction and preliminary screening. The right panel shows the flowchart of computational high-throughput screening of photoelectric materials.

Choosing appropriate features (descriptors) is known to influence the accuracy of the machine-learning algorithm more than the algorithm itself according to many cheminformatics investigations.[40, 41] The stability, band gap, and the formation heat of perovskites are well described by the octahedral factor[42, 43], $Of = \frac{r_A}{r_B}$, where $r_A$ is the ion radii of the octahedral central atom, and $r_B$ is the ion radii of the octahedral vertex atom. In consideration of the distortion and stacking of octahedron in OOHs (see Figure 1(b)), we have tried to adopt the $Of$ as a feature. However, we also propose distorted stacked octahedral factors ($DSOf_n$, $n = 1\sim8$) as a modification. For example, one of them is $DSOf_5 = \frac{r_A + r_{B1}}{r_{B1} + r_{B2}}$, where $r$ is the ion radii of specific atoms *A*, *B1*, or *B2* within the OOHs. (See Table S1 for the definition of all features). We consider $DSOf$ as better features than $Of$ since $DSOf$ can describe the geometric relation of the inequitable atoms in octahedrons in an even better fashion. To verify, we compare four different feature vectors (features of element-related properties; features of element-related properties plus $Of$; features of element-related properties plus $DSOf$; features of element-related properties plus both $DSOf$ and $Of$), and find that the feature vectors including element-related properties and $DSOf$ give the best prediction accuracy among the others (see Table S2).

We have tested various algorithms such as the support vector regression (SVR),[44] random forest regression (RFR),[45] bagging[46] and gradient boosting regression (GBR), which all have been successfully used for machine-learning prediction in materials.[47] The results of 10-fold cross-validation regression analysis and grid searching best hyper-parameters technique show that the GBR model outperforms others and is the best model with the lowest mean square error (MSE = 0.086) and largest coefficient of determination value ($R^2$= 0.835) (See Figure 3(a) and Figure S2). The subplot in Figure 3(a) clearly shows that the MSE deviances of training and test learning curve decline gradually with the increase of the data size, certifying that the problem of over-fitting does not appear. These MSE and $R^2$ values are also comparable to those in the machine-learning prediction of other materials.[10, 11, 43] Thus, the algorithm we adopted is reliable for further analysis.

The best approach to avoid an overly complex model is to choose a suitable number of features that could perfectly reflect materials' properties. In our present study, we choose 62 initial features, including the aforementioned $DSOf$ and other features of element-related properties, to describe the details of OOHs in the chemical space collectively. Using the principal component analysis (PCA) method[48], we acquire the importance of all features, and then we perform a feature reduction by excluding the features with less impact until $R^2$ achieves a maximum (see Figure S1). In the end, we find that 26 features form an optimal vector of features including eight $DSOf$ and the 18 element-related properties of $A$, $B1$, $B2$, $X1$ and $X2$. The 18 element-related properties are the ionization energies, $IE_A$, $IE_{B1}$, $IE_{B2}$, the ionic polarizabilities, $P_A$, $P_{X1}$, $P_{B2}$, the electronegativities (Martynov-Batsanov[49] scale), $X_A$, $X_{B1}$, $X_{B2}$, the p-orbital radii for $B1$ and $X2$, $r_{B1(p)}$, $r_{X2(p)}$, the inner and outer valence electron orbital radii for $A$, $r_{A(in)}$ and $r_{A(out)}$, the numbers of inner or outer valence electrons for $A$, $n_{A(in)}$ and $n_{A(out)}$, the ionic charge for $A$, $IC_A$, the ion radii for $A$, $r_A$, and electron affinity for $B1$, $EA_{B1}$. (see Table S1 for feature details.) As illustrated in Figure 2(a), the top features, i.e., $DSOf_1$, $DSOf_3$, $DSOf_5$ and $DSOf_7$ (be short for 'top four $DSOf$')), play the most essential roles in determining the OOHs' band gaps, followed by $IE_A$, $r_{A(in)}$ and $X_A$, i.e., the ionization energy, the inner valence electrons orbital radii and electronegativities for $A$-sites, which also have very important effects. Note that other four $DSOf$ features, i.e., $DSOf_2$, $DSOf_4$, $DSOf_6$, and $DSOf_8$ (be short for 'minor four $DSOf$') are of minor importance. Interestingly, the top four $DSOf$ all contain $r_A$ (ion radii of $A$), while the minor four $DSOf$ do not contain $A$-related properties. Thus, it appears that the OOHs' band gaps are more sensitive to the element−related properties of the $A$-site (octahedron center) than those of other sites. This behavior is similar to what was observed in perovskites.[14] Figure 2(b) shows a heat map of the Pearson correlation coefficients between the features. Strong correlations among top four $DSOf$ indicate that these features are highly coupled. However, weak correlations between top four $DSOf$ features and the minor four $DSOf$ features have been observed. In addition, the strong correlations between top four $DSOf$ and $r_A$ indicate that these features are highly coupled.

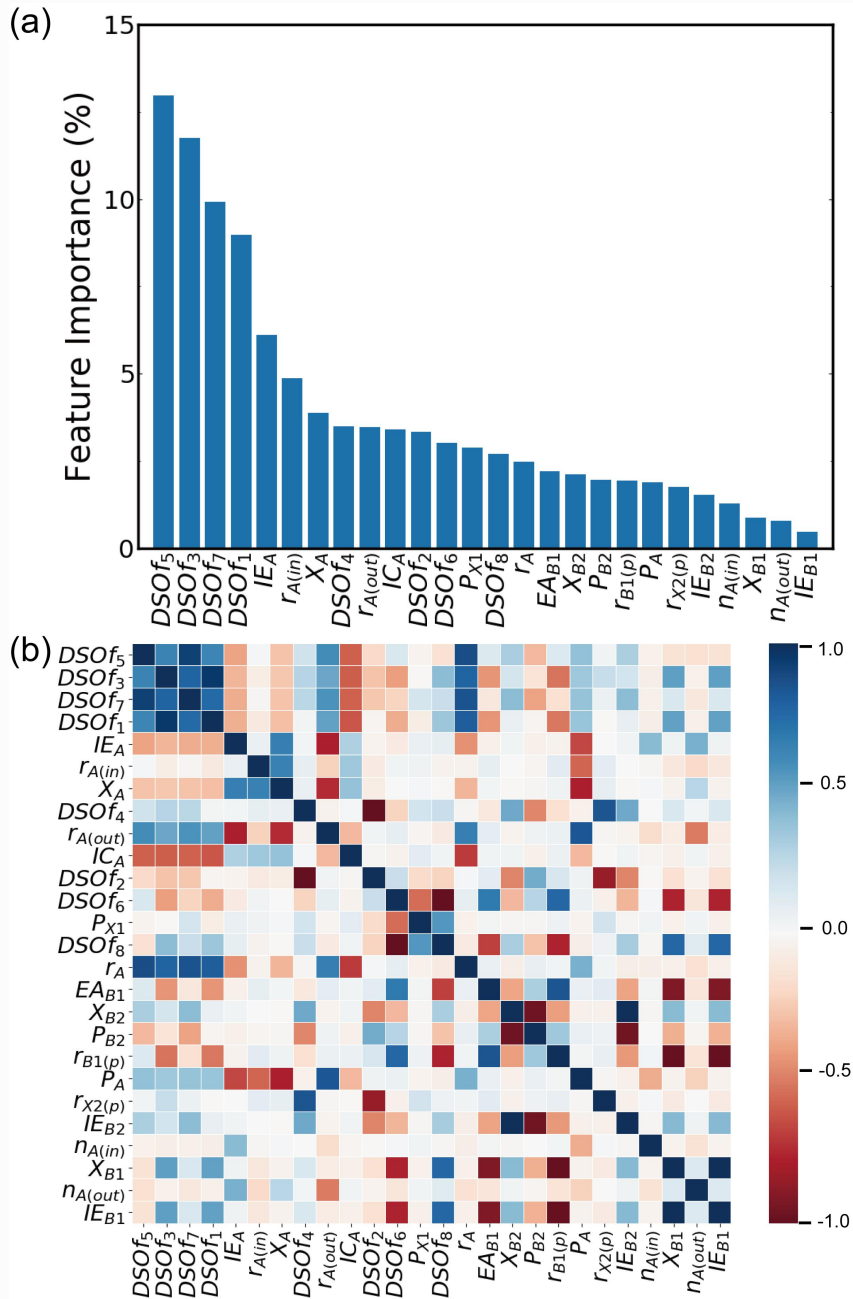

Figure 2. Importance and correlation of 26 optimal features. (a) features are ranked with importance using GBR algorithm, and (b) Statistical heat map shows the correlation of these features.

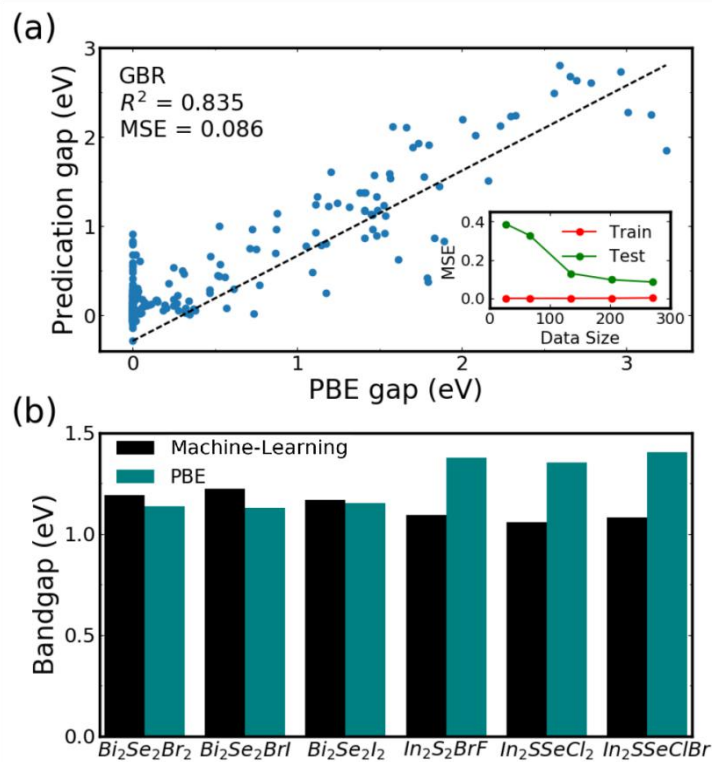

Figure 3. The comparison between the band gaps predicted by GBR model and DFT calculations. (a) The fitting results of the band gaps between DFT-calculated values and the ten-fold cross-validation predicted ones for 300 OOHs. The reliability of the machine-learning model is evaluated by the coefficient of determination ($R^2$) and the mean squared error (MSE). The inset indicates the convergence of learning curve for the ten-fold cross-validation split of the data. (b) A comparison between machine-learning-predicted and PBE-calculated band gaps of six OOHs.

After determining the importance and correlations of the primary features, we apply machine-learning algorithms to predict the band gaps of the remaining 5000 OOHs. As displayed in Figure 3(b), the comparison between machine-learning-predicted and DFT-calculated band gaps of six selected OOHs shows that the error is reasonable and the MSE for preliminary screened 73 OOHs is about 0.071, suggesting that the current machine-learning model is quite efficient. In addition, the error of mixing S/Se, Br/F and Cl/Br OOHs is larger than other three OOHs in Figure 3(b). The error of machine-learning prediction depends on the size of data. The combination of mixing S/Se, Br/F and Cl/Br will generate a large number of different OOHs, which may be four times more than those of the non-mixing OOHs.

The mixing also brings a lower symmetry and makes the structures diverse. Therefore, the mixing S/Se, Br/F and Cl/Br of OOHs needs larger data sets than non-mixing OOHs. More importantly, the time savings is order of magnitudes when comparing the time for performing the DFT calculation and applying our machine-learning algorithm. Therefore, the present machine-learning-throughput strategy provides a possibility for searching for complex systems like OOHs with reasonable accuracy and accelerated speed.

Here we focus on three OOHs ($Bi_2Se_2Br_2$, $Bi_2Se_2BrI$, and $Bi_2Se_2I_2$) with promising optoelectronic properties. The geometric structures of $Bi_2Se_2X_2$ (X=Br/I) and $Bi_2Se_2BrI$ are displayed in Figure S3, and their structural parameters are listed in Table S4, where their space groups are *Pmmn* and *Pmm2* (Janus $Bi_2Se_2BrI$ results in the reduction of symmetry), respectively. The phonon dispersions of these structures are calculated (as given in Figure S7 (a), (c) and (e)), showing these OOHs are dynamically stable. In addition, *ab initio* molecular dynamics simulations are also performed at room temperature. The time-dependent variations of energy per atom are oscillating within a very narrow range (Figure S7 (b), (d), and (f)), manifesting that these structures can maintain their structures under thermal disturbance.

Their electronic properties are systematically investigated. 2D $Bi_2Se_2Br_2$, $Bi_2Se_2BrI$, and $Bi_2Se_2I_2$ are found to be semiconducting with indirect band gaps of 1.137, 1.130 and 1.151 eV (Figure 4a, c, e) at Perdew-Burke-Ernzerh (PBE) functional level,[50] respectively. While at the level of HSE06 functional[51] and in consideration with the spin-orbital coupling (SOC) effect, the band gaps slightly increase to 1.371, 1.218 and 1.233 eV (see Figure S5), respectively, indicating that these band gaps are suitable for photovoltaic devices. As shown in Figure 4(a), (c) and (e), by analyzing the dispersion of the conduction band minima (CBM) and the valence band maxima (VBM), we find that the electron effective mass along S-X (*x* direction in real space) is lower than that of S-Y (*y* direction in real space), implying a directional anisotropic electron transport property. In addition, the electron effective masses are lower than those of holes in these three materials, indicating the asymmetry of the charge carriers.

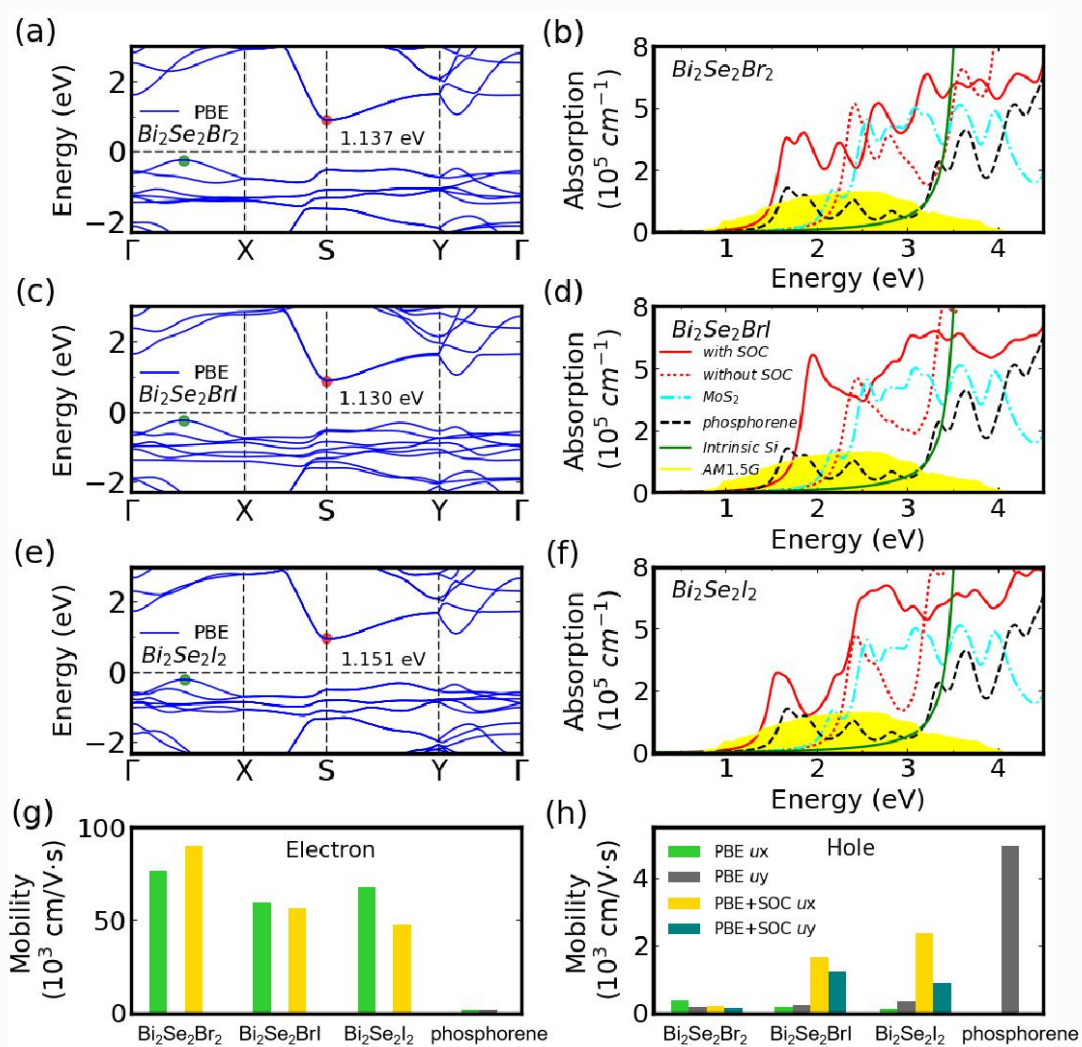

Figure 4. Calculated electronic band structures, mobilities and absorption spectra of three OOHs: $Bi_2Se_2Br_2$, $Bi_2Se_2BrI$ and $Bi_2Se_2I_2$. (a)(c)(e) Band Structures at PBE level. (b)(d)(f) Absorption spectra at $G_0W_0$+BSE level with/without SOC. Adsorption spectra of monolayer $MoS_2$, phosphorene and intrinsic Si are also included for comparison. The yellow area indicates the incident AM1.5G solar flux. Note green solid lines present the adsorption spectra of intrinsic Si. (g) Electron mobility and (h) hole mobility along $x$ direction and $y$ direction at PBE and PBE+SOC level. Carrier mobilities of phosphorene monolayer is also included for comparison. Note that in phosphorene the SOC is too weak to be considered.

The carrier (electron or hole) mobility has a great impact on the performance of electronic devices. We have calculated the carrier mobilities and the details can be found in the supplementary information (See Figure S9-S12, Table S5, and S6. Figure 4(g) and (h)). The results of carrier mobility for 2D $Bi_2Se_2Br_2$, $Bi_2Se_2BrI$, and $Bi_2Se_2I_2$, reveal that electron mobility is much larger than that of the holes, indicating the charge/hole asymmetry, which is dominated by the carrier effective mass asymmetry. In Figure 4(g), one may see that the electron mobility along *x* direction is several orders of magnitude larger than that along *y* direction, suggesting a very directional anisotropy for the motion of electrons, which are mainly attributed to the direction-dependent deformation potentials and effective electron masses. The electron mobility of $Bi_2Se_2Br_2$ can be as ultrahigh as $7.66 \times 10^4$ cm$^2$V$^{-1}$s$^{-1}$ along *x* direction, and those of $Bi_2Se_2I_2$ and $Bi_2Se_2BrI$ have lower values in the same order of magnitude (Figure 4(g) and Table S4). It is interesting to note that the mobility of these materials are much larger than that of phosphorene ($5 \times 10^3$ cm$^2$V$^{-1}$s$^{-1}$) and SnSe monolayers ($1 \times 10^4$ cm$^2$V$^{-1}$s$^{-1}$) by the same calculation method,[52, 53] indicating that these OOHs would also be compelling candidates for potential applications in future high-speed electronic devices, in particular in resistive random access memory devices. The carrier mobilities of above three OOHs are also calculated at PBE+SOC level. Figure 4(g) shows that the electron mobility does not alter remarkably because the SOC results in only a smaller electron deformation potential and makes the electron effective mass heavier (see Table S5, S6). In Figure 4h, by comparing with that of $Bi_2Se_2Br_2$, we find that the great improvement of hole mobility of $Bi_2Se_2I_2$ and $Bi_2Se_2BrI$ (including both heavy elements Bi and I) are mainly attributed by the drastic reduction of hole effective masses (see Table S5, S6), which may be induced by the strong SOC effect. In addition, due to the limitation of longitudinal optical phonons on the carrier mobility,[54] their experimental mobilities should be lower than the theoretical values.

Figure 4(b), (d), and (f) present the absorption coefficients of these OOHs, where those of intrinsic Si, phosphorene, and $MoS_2$, and the incident AM1.5G solar spectrum[55] are also included for comparison. It should be mentioned that our calculated absorption coefficients for Si and phosphorene are in accordance with the previous results.[56, 57] The three OOHs have remarkably higher absorbance coefficients ($10^5$ cm$^{-1}$) than those of Si, phosphorene, and $MoS_2$. Furthermore, their absorptions cover almost the entire incident solar spectrum, which could guarantee high conversion efficiency in terms of utilization of solar energy. Optical absorption is closely related to electronic structures. If the SOC effect has a great effect on the electronic structure, the calculation of optical absorption should need to consider the SOC effect. The SOC effect has a pronounced effect on the electronic structure of them (Figure S5). Therefore, the absorption of them is performed by $G_0W_0$+BSE with SOC effect. The obvious shift of the absorption edges of these OOHs toward low energy direction is primarily owing to the SOC effect, leading to wider absorption windows. The improvements of the absorption coefficients of $Bi_2Se_2I_2$ and $Bi_2Se_2BrI$ benefit from the increasing of the valley numbers of conduction bands and valence bands due to strong SOC effect, which also enhances the electron transitions from the valence band to the conduction band (see Figure S4-S5). The absorption of Janus $Bi_2Se_2BrI$ is greater than those of two others. Their outstanding absorbance coefficients and wide absorption range make these OOHs potential materials for photovoltaic solar cells and optoelectronic devices. Interestingly, there are some other bismuth-based oxyhalides that have the similar chemical formula,[58, 59] but their structures are distinctly different from OOHs, and thus our results are totally new findings.

In summary, we have developed a method combining the high-throughput *ab initio* calculations with machine learning to discover optoelectronic OOHs with good performance. Six stable OOHs with proper band gaps and high carrier mobilities have been successfully screened out from 5000 OOHs. Specifically, three of them, say, $Bi_2Se_2Br_2$, $Bi_2Se_2BrI$, and $Bi_2Se_2I_2$, are predicted to have pronounced absorption capacities in the whole range of the solar spectrum, which would be promising candidates for potential applications in electronic and photovoltaic devices. In conventional high-throughput calculations, the whole chemical space should be searched at the DFT level, while our present work discloses that the incorporation of the usual high-throughput calculations with machine-learning technology can greatly reduce the searching space, which could accelerate the discovery of novel materials with desired properties. Moreover, the distorted stacked octahedral factor features we proposed here are also confirmed to be conducive to construct structural descriptors of other complex materials, and are helpful for the establishment of the machine-learning model. The above approach could overcome a major obstruction in the traditional trial-error method and is applicable to the discovery and design of novel functional materials.

## ASSOCIATED CONTENT

**Supporting information Available:**

The computational methods, gradient boosted regression, model evaluation, initial features with definition, feature reduction, algorithm selection, comparison with various feature combinations, comparison between Machine-learning-predicted and DFT-calculated band gaps, structural details, electronic structures, phonon dispersions, AIMD evolutions, and carrier mobility. (PDF)

## AUTHOR INFORMATION

**Correspondence Author**

*E-mail: yan@ucas.ac.cn

*E-mail: gsu@ucas.ac.cn

**ORCID**

James P. Lewis: **0000-0002-6724-3483**

Qing-Bo Yan: **0000-0002-1001-1390**

Gang Su: **0000-0002-8149-4342**

**Notes**

The authors declare no competing financial interest.

## ACKNOWLEDGMENTS

This work is supported in part by the National Key R&D Program of China (Grant No. 2018YFA0305800), the Strategic Priority Research Program of CAS (Grant No. XDB28000000), the NSFC (Grant No. 11834014), Beijing Municipal Science and Technology Commission (Grant No. Z118100004218001) and JPL received support from the Thousand Talent program of the Chinese Academy of Sciences. The calculations were performed on Era at the Supercomputing Center of Chinese